\documentstyle[aps,preprint]{revtex}
\begin{document}
\draft

\title{Creation of spin 1/2 particles by an electric field in de Sitter space}

\author{ V\'{\i}ctor M. Villalba\footnote{e-mail address:
villalba@dino.conicit.ve}}

\address{Centro de F\'{\i}sica, Instituto Venezolano de Investigaciones
Cient\'{\i}ficas \\ IVIC, Apdo. 21827, Caracas 1020-A Venezuela}

\address{Centre de Physique Th\'eorique C.N.R.S. \\ Luminy- Case 907-F-13288
Marseille Cedex 9, France}

\maketitle

\begin{abstract}
In the present article we solve the Dirac equation in a de Sitter
universe when a constant electric field is present. Using the Bogoliubov
transformations, we compute the rate of spin 1/2  created particles by the
electric field. We compare our results with the scalar case. We also analyze
the behavior of the density of particles created in the limit H=0, when de
Sitter background reduces to a flat space-time.
\end{abstract}

During the last two decades, quantum field theory in curved spaces has been
extensively discussed, mainly in the study of particle creation in vicinity
of black holes as well as in expanding cosmological universes. Here we are
talking about a regime where the spacetime curvature remains significantly
less than the Planck scale, and therefore one may usefully employ the
semiclassical approximation in which the gravitational field is treated
classically and we analyze quantized elementary particles propagating on
this classical background. This is perhaps the simplest theory that one can
use for discussing quantum effects in the presence of gravitational fields
in the early universe. In this framework we are interested in discussing
spin 1/2 particle creation by a constant electric field in a (1+1) de Sitter
space. This problem has been preceded by a series of articles by Lotze \cite
{Lotze1,Lotze2} where the author analyzes the influence of interacting
fields on the rate of particles produced by the gravity. In particular, he
considers the electromagnetic interaction $ie\bar \Psi \gamma ^\mu A_\mu
\Psi $ between the electron-positron field and the potential $A_\mu $ in
expanding spatially flat Robertson-Walker Universes, and more recently,the
problem of pair production by an electric field in (1+1)-dimensional de
Sitter space has been discussed for the case of scalar particles \cite
{garri2}. Here, the idea was to establish a comparison with the results
obtained after computing the rate of membranes created by an anti symmetric
tensor field\cite{garri1,Brown} In the present article, as a natural
extension of the results reported in Ref. \cite{garri2} we solve the Dirac
equation in (1+1) de Sitter universe when a constant electric field is
present. After identifying the the ``in'' and ``out'' vacuum states, we
compute, via the Bogoliubov transformations, the rate of particles created
by the electric and gravitational fields. We compare our results with those
obtained in the scalar case.

The line element associated with a (1+1) de Sitter universe takes the simple
form
\begin{equation}
\label{1}ds^2=-dt^2+e^{2Ht}dx^2
\end{equation}
which, in terms of the conformal time $\eta =-\frac 1He^{-Ht}$ reduces to
\begin{equation}
\label{2}\frac 1{H^2\eta ^2}(-d\eta ^2+dx^2)
\end{equation}
Since we are interested in analyzing Dirac particle production by a constant
electric field in a de Sitter background, in order to establish a comparison
with the pure gravitational contribution, it is advisable to have as a
previous result the rate of spin 1/2 pair production in de (1+1) Sitter
universe. For this purpose we proceed to solve the Dirac equation in the
(1+1) de Sitter space (\ref{2}). The covariant generalization of the Dirac
equation reads
\begin{equation}
\label{Dirac}\left( \gamma ^\alpha (\partial _\alpha -\Gamma _\alpha
)+m\right) \Psi =0
\end{equation}
in the background field where $\gamma ^\alpha $ are the curved Dirac
matrices satisfying the commutation relation \{$\gamma ^\alpha ,\gamma
^\beta \}=2g^{\alpha \beta },$ $\Gamma _\alpha $ are the spinor connections,
which for a diagonal metric as (\ref{1}) reduces to
\begin{equation}
\label{con}\Gamma _\lambda =-\frac 14g_{\mu \alpha }\Gamma _{\nu \lambda
}^\alpha s^{\mu \nu },\
\end{equation}
where
\begin{equation}
s^{\mu \nu }=\frac 12(\gamma ^\mu \gamma ^\nu -\gamma ^\nu \gamma ^\mu )
\end{equation}
Choosing to work in the diagonal tetrad gauge, we have that the gamma
matrices take the form
\begin{equation}
\label{matr}\gamma ^0=-H\eta \bar \gamma ^0,\ \gamma ^1=-H\eta \bar \gamma
^1
\end{equation}
where the flat gamma matrices $\bar \gamma ^\alpha $ satisfy \{$\bar \gamma
^\alpha ,\bar \gamma ^\beta \}=2\eta ^{\alpha \beta }.$ Then, from (\ref{con}%
), we obtain that the only non zero connection $\Gamma _\alpha $ reads
\begin{equation}
\label{gamma1}\Gamma _1=-\frac 1{2\eta }\bar \gamma ^0\bar \gamma ^1
\end{equation}
Since we are dealing with a (1+1) dimension space, we have that gamma
matrices can be represented in terms of the Pauli matrices. In particular,
we can make the following election:
\begin{equation}
\label{repr}\bar \gamma ^0=i\sigma ^3,\ \bar \gamma ^1=\sigma ^1
\end{equation}
substituting the gamma matrices (\ref{matr}) as well as the spinor
connection (\ref{gamma1}) into the Dirac equation (\ref{Dirac}), and
introducing the spinor $\Phi $ related to $\Psi $ as follows
\begin{equation}
\label{Fi}\Phi =\eta ^{-1/2}\Psi
\end{equation}
we obtain that the problem of solving Eq. (\ref{Dirac}) reduces to find
solutions of the coupled system of equations
\begin{equation}
\label{odin}\left( \frac d{d\eta }+\frac{im}{H\eta }\right) \Phi _1+k_x\Phi
_2=0
\end{equation}
\begin{equation}
\label{dva}\left( -\frac d{d\eta }+\frac{im}{H\eta }\right) \Phi _2+k_x\Phi
_1=0
\end{equation}
where $k_x$ is the eigenvalue of the linear momentum operator $-i\partial _x$
consequently, we can write down the spinor $\Phi $ as $\Phi =\Phi _0(\eta
)e^{ik_xx}$ Substituting (\ref{dva}) into (\ref{odin}) and vice-versa, we
arrive at
\begin{equation}
\label{second}\left( \frac{d^2}{d\eta ^2}+(\frac{m^2}{H^2}\mp \frac{im}H)%
\frac 1{\eta ^2}+k_x^2\right) \left(
\begin{array}{c}
\Phi _1 \\
\Phi _2
\end{array}
\right) =0
\end{equation}
The solutions of the system of equations (\ref{second}) can be expressed in
terms of cylinder functions $Z_\nu (z)$ as follows
\begin{equation}
\label{sol}\Phi _1=\eta ^{1/2}Z_\nu (k_x\eta ),\ \Phi _2=-\eta ^{1/2}Z_{\nu
-1}(k_x\eta )
\end{equation}
where the parameter $\nu $ is given by the expression
\begin{equation}
\nu =\frac 12+\frac{im}H
\end{equation}
In order to analyze quantum effects in the background field (\ref{2}) we
have to define vacuum states in the asymptotic regions. The notion of
positive and negative frequency states is associated with the existence of a
timelike Killing vector. In expanding cosmological Universes that vector
does not exist, and we have to appeal to another methods for obtaining a
reasonable definition of the vacuum state. Among the different approaches we
have that the so-called adiabatic method is perhaps the most popular. \cite
{Parker1,birrel} In our particular case, since the asymptotic behavior for
scalar fields in the background field (\ref{2}) was discussed in Ref. \cite
{garri2}, we already know the asymptotic form that our in and out vacua
should take. Nevertheless, a natural criterion for choosing the initial and
final states is to look at the behavior of the quasiclassical modes
described by the Hamilton-Jacobi equation
\begin{equation}
\label{HJ}g^{\alpha \beta }S_{,\alpha }S_{,\beta }+m^2=0
\end{equation}
which in the metric (\ref{2}) can be written as a sum
\begin{equation}
S=T(\eta )+F(x)
\end{equation}
and, after introducing the constant of separation $k$ associated with the
space variable $x$, we arrive at the first order differential equation:
\begin{equation}
\label{HJS}k^2+\frac{m^2}{H^2\eta ^2}=(S_{,\eta })^2
\end{equation}
Then the quasiclassical behavior of the solution of the Dirac equation (\ref
{Dirac}) in the metric (\ref{2}) is
\begin{equation}
\Psi \rightarrow \Psi _0(x)\exp (\pm i\int \sqrt{k^2+\frac{m^2}{H^2\eta ^2}}%
d\eta )=\Psi _0(x)\frac{\exp (\pm ik\sqrt{\eta ^2+\frac{m^2}{k^2H^2}}\eta
^{\pm im/H}}{\left( \frac m{kH}+\sqrt{\eta ^2+\frac{m^2}{k^2H^2}}\right)
^{im/H}}
\end{equation}
then, as $\eta \rightarrow -\infty $ ($t\rightarrow \infty )$ we find
\begin{equation}
\label{u}\Psi \rightarrow \Psi _0(x)\exp (\mp ik\eta )
\end{equation}
and as $\eta \rightarrow 0$ ($t\rightarrow \infty )$
\begin{equation}
\label{v}\Psi \rightarrow \Psi _0(x)(\frac{2m}{Hk})^{-im/H}\eta ^{\pm
im/H}=\Psi _0(x)(\frac{2m}{Hk})^{-im/H}e^{\mp imt/2}
\end{equation}
where the upper and lower signs in (\ref{u}) and (\ref{v}) are associated
with positive and negative frequencies respectively. Then, looking at the
asymptotic behavior of the Hankel functions,
\begin{equation}
H_\nu ^{(2)}(x)\approx \sqrt{\frac 2{\pi x}}e^{-i(x-\nu \pi /2-\pi /4)}\
x\rightarrow \infty
\end{equation}
we have that the ``in'' state as $\eta \rightarrow -\infty $ can be
expressed in terms of $H_\nu ^{(2)}(x)$ as
\begin{equation}
\label{H1}\Phi _1=\eta ^{1/2}H_\nu ^{(2)}(k_x\eta ),
\end{equation}
substituting (\ref{H1}) into (\ref{d1}) we readily find that $\Phi _2=-\eta
^{1/2}H_{\nu -1}^{(2)}(k_x\eta ).$ The expression (\ref{H1}) corresponds to
the truly de Sitter invariant vacuum as it was pointed out by Bunch and
Davies \cite{Bunch}. From the definition of $H_\nu ^{(2)}(z)$ in terms of
the Bessel function $J_\nu (z)$
\begin{equation}
H_\nu ^{(2)}(z)=\frac{e^{i\pi \nu }J_\nu (z)-J_{-\nu }(z)}{i\sin \pi \nu }
\end{equation}
and the behavior for small values of $z$
\begin{equation}
J_\nu (z)\rightarrow \frac{z^\nu }{2^\nu \Gamma (\nu +1)},\
\end{equation}
and as $\eta \rightarrow 0$ we have that in the vicinity of $\eta =0$ the
spinor component $\Phi _1$ takes the form
\begin{equation}
\Phi _1\rightarrow \frac{\eta ^{1/2}}{i\pi \nu }\left( \frac{(k_x\eta )^\nu
}{2^\nu }e^{i\pi \nu }\Gamma (1-\nu )-2^\nu \Gamma (\nu +1)(k_x\eta )^{-\nu
}\right)
\end{equation}
then, looking at the quasiclassical asymptotic behavior of the ``out''
states (\ref{v}) and the linear decomposition between positive and negative
frequency modes
\begin{equation}
\label{modes}\Phi _{in,k}^{+}=\alpha _k\Phi _{out,k}^{+}+\beta _k\Phi
_{out,k}^{-}
\end{equation}
we arrive at
\begin{equation}
\label{thermal}\mid \alpha _k\mid ^2=e^{2\pi m/H}\mid \beta _k\mid ^2
\end{equation}
following Mishima and Nakayama \cite{Mishima}, we can already establish that
the spectrum of particles created is a thermal one with temperature $T=H/{%
2\pi }$ From $\mid \alpha _k\mid ^2+\mid \beta _k\mid ^2=1$, one readily gets

\begin{equation}
\mid \beta _k\mid ^2=\frac 1{e^{2\pi m/H}+1}
\end{equation}
then, we obtain a thermal Fermi-Dirac distribution for $\mid \beta _k\mid ^2$%
, which for $m>>H$ gives as result that the number of particles per unit
coordinate volume is $\frac{d{\cal N}}{dx} \approx e^{-2\pi m/H}\frac{dk}{%
2\pi}$. Notice that for the Dirac case, we do not need to impose any
restriction on the relation between $m$ and $H$ in order to obtain well
defined ``out'' states.

Let us extend our problem to the case when a constant electric field
minimally coupled to the spinor field is present. Then the Dirac equation
can be written as
\begin{equation}
\label{DA}\left( \gamma ^\alpha (\partial _\alpha -\Gamma _\alpha
)+m-ie\gamma ^\alpha A_\alpha \right) \Psi =0
\end{equation}
where the vector potential $A_\mu $ associated with a constant electric
field $E_0$ can be written in the comoving time $\eta $ as follows
\begin{equation}
\label{A}A_0=0,\ A_1=-\frac{E_0}He^{Ht}=\frac{E_0}{H^2\eta }
\end{equation}
substituting (\ref{A}) and $\Gamma _\alpha $ given by (\ref{gamma1}) into (%
\ref{DA}) we find
\begin{equation}
\label{DC}\left(\bar\gamma ^0\partial _\eta +\bar\gamma ^1\partial _x-\frac
m{H\eta
}-\frac{ieE_0}{H^2\eta }\bar\gamma ^1\right) \Phi =0
\end{equation}
using the gamma's representation (\ref{repr}) we find that eq. (\ref{DC})
yields
\begin{equation}
\label{d1}\left( \frac d{d\eta }+\frac{im}{H\eta }\right) \Phi _1+\left( k_x-%
\frac{eE_0}{H^2\eta }\right) \Phi _2=0
\end{equation}
\begin{equation}
\label{d2}\left( -\frac d{d\eta }+\frac{im}{H\eta }\right) \Phi _2+\left(
k_x-\frac{eE_0}{H^2\eta }\right) \Phi _1=0
\end{equation}
the system of equations (\ref{d1})-(\ref{d2}) can be written in the matrix
form
\begin{equation}
\label{Fai}\left[ \frac d{d\eta }+\frac{im}{H\eta }\sigma _3+i\sigma
_2\left( k_x-\frac{eE_0}{H^2\eta }\right) \right] \left(
\begin{array}{c}
\Phi_1 \\
\Phi_2
\end{array}
\right) =0
\end{equation}
and with the help of the similarity transformation $T$, acting on the Pauli
matrices $\sigma _i$ as $T\sigma _iT^{-1}=\sigma _i^{\prime }$
\begin{equation}
T=\left(
\begin{array}{cc}
1 & -i \\
-i & 1
\end{array}
\right) =(1-i\sigma _1)
\end{equation}
and the auxiliary spinor $\Theta =\eta ^{1/2}\Phi ,$ we reduce the coupled
system of equations (\ref{d1})-(\ref{d2}) to the Whittaker equations
\begin{equation}
\label{wi}\left( \frac{d^2}{d\rho ^2}-\frac 14+(\frac{ieE_0}{H^2}\pm \frac 12%
)\frac 1\rho +\frac{+e^2E_0^2/H^4+m^2/H^2+1/4}{\rho ^2}\right) \left(
\begin{array}{c}
\Theta _1 \\
\Theta _2
\end{array}
\right) =0
\end{equation}
having as a solution
\begin{equation}
\label{Ml}M_{\lambda ,\mu }(\rho )=\Theta _{1,2}=\eta ^{1/2}\Phi
_{1,2}=c_{1,2}\rho ^{\mu +1/2}e^{-\rho /2}M(\mu -\lambda +\frac 12,2\mu
+1,\rho )
\end{equation}
where we have introduced the auxiliary variable $\rho =2ik_x\eta $, and the
parameters $\mu $ and $\lambda $ read
\begin{equation}
\label{lambda}\mu =\frac 1H\sqrt{-e^2E_0^2/H^2-m^2}=i\mid \mu \mid ,\
\lambda =\pm \frac 12+\frac{ieE_0}{H^2}
\end{equation}
where the upper and lower signs in (\ref{lambda}) and (\ref{Ml}) correspond
to $\Phi _1$ and $\Phi _2$ respectively. A second set of solutions of the
Eq. (\ref{wi}) can be expressed in terms of the Whittaker functions $%
W_{\lambda ,\mu }(z)$ \cite{Gradshteyn}
\begin{equation}
\ W_{\lambda ,\mu }(z)=z^{\mu +1/2}e^{-z/2}U(\frac 12-\lambda +\mu ,2\mu
+1,z)
\end{equation}
Using the quasiclassical asymptotae, we can choose positive and negative
frequency modes. In fact, looking at the asymptotic behavior for large
values of $\mid z\mid $
\begin{equation}
W_{\lambda ,\mu }(z)\rightarrow e^{-z/2}z^\lambda
\end{equation}
and for small $z$%
\begin{equation}
M_{\lambda ,\mu }(z)\rightarrow z^{\mu +\frac 12}
\end{equation}
we obtain that $\Phi _{in,k}^{+}=cW_{\lambda ,\mu }(z)$, where $c$ is a
constant of normalization, and $\Phi _{out,k}^{+}\sim M_{\lambda ,\mu }(z)$.
Then, using the relation \cite{Gradshteyn}
\begin{equation}
\label{W}M_{\lambda ,\mu }(z)=\Gamma (2\mu +1)e^{i\pi \lambda }\left[ \frac{%
W_{-\lambda ,\mu }(e^{i\pi }z)}{\Gamma (\mu -\lambda +\frac 12)}+\frac{%
W_{\lambda ,\mu }(z)}{\Gamma (\mu +\lambda +\frac 12)}\exp \left[ i\pi
(\lambda -\mu -\frac 12)\right] \right]
\end{equation}
\begin{equation}
-\frac{3\pi }2<\arg z<\frac \pi 2;\ 2\mu \neq -1,-2...
\end{equation}
from Eq. (\ref{W}), and the relation \cite{Gradshteyn}
\begin{equation}
\mid \Gamma (iy)\mid ^2=\frac \pi {y\sinh \pi y}
\end{equation}
we arrive at
\begin{equation}
\label{espe}\frac{\mid \alpha _k\mid ^2}{\mid \beta _k\mid ^2}=\frac{\mid
\sinh \pi (\mid \mu \mid -eE_0/H^2)\mid e^{2\pi( \mid \mu \mid
-eE_0/H^2)}}{\mid \sinh
\pi (\mid \mu \mid +eE_0/H^2)\mid }
\end{equation}
the expression (\ref{espe}) takes simpler form when $\mid \mu \pm
eE_0/H^2\mid >>1$ Taking into account the relation $\mid \alpha _k\mid
^2+\mid \beta _k\mid ^2=1$ we finally obtain the thermal distribution
\begin{equation}
\label{spec}\mid \beta _k\mid ^2=\frac 1{1+e^{2\pi \mid \mu \mid -2e\pi
E_0/H^2}}\approx e^{(-\frac{2\pi }{H^2}\left[
(e^2E_0^2+m^2H^2)^{1/2}-eE_0\right] )}
\end{equation}
which shows that the temperature depends on the intensity of the electric
field $E_0$. Also we have, as for the scalar case, that $\mid \beta _k\mid
^2 $ changes according the sign of $k_x$. In fact, from (\ref{d1})- (\ref{d2}%
), it becomes clear that changing the sign of $k_x$ is equivalent to replace
$e$ by $-e$ in (\ref{spec}).

It is interesting to analyze the form expression (\ref{spec}) takes  in the
limit $%
H\rightarrow 0.$ In this case we have that the line element (\ref{1}) reduces
to a
2-d Minkowski (flat) metric.

Since the exponent of $\mid \beta _k\mid ^2$ in (\ref{spec}) can be rewritten
as:
\begin{equation}
\frac{-2\pi }{H^2}\left[ (e^2E_0^2+m^2H^2)^{1/2}-eE_0\right] =\frac{-2\pi
eE_0}{H^2}\left[ (1+\frac{m^2H^2}{e^2E_0^2})^{1/2}-1\right]
\end{equation}
and in the limit  when   $H$ goes to zero,  we have
\begin{equation}
\lim _{H\rightarrow 0}[(1+\frac{m^2H^2}{e^2E_0^2})^{1/2}-1]\rightarrow \frac{%
m^2H^2}{2e^2E_0^2}
\end{equation}
we readily obtain that in the flat limit  the density of particles created
reduces to
\begin{equation}
\label{flat}\mid \beta _k\mid _{{\rm flat}}^2=\lim _{H\rightarrow 0}\mid
\beta _k\mid ^2=e^{-\pi m^2/(eE_0)}
\end{equation}
which is the result obtained by Narozhny and Nikishov
\cite{nikishov1,nikishov2}
in analyzing the phenomenon of pair creation by a strong constant electric
field when the transverse linear momentum $p_{\perp}$ is equated to zero. It
is worth mentioning that expression (\ref{flat}) is proportional to the first
term of the series obtained by Schwinger \cite{schwinger}, giving the
probability,
per unit time, and per unit volume, that a pair is created by the constant
electric
field.

\acknowledgments

\noindent The author wishes to express his gratitude to the Centre de
Physique Th\'eorique for its hospitality. Also the author wishes to thank
Bruce Bassett for reading and improving the manuscript as well as the referee
for drawing my attention to the Schwinger's result. This work was partially
financed by the CONICIT of Venezuela and the Fundaci\'on Polar.

\end{document}